\documentclass[%
superscriptaddress,
showpacs,preprintnumbers,
nofootinbib,
 amsmath,amssymb,
 aps,
 twocolumn,
prl,
nofootinbib
]{revtex4}
\usepackage{braket}
\usepackage{bm}
\usepackage{dcolumn}
\usepackage[dvipdfmx,final]{graphicx}
\input{colordvi.tex}
\usepackage{longtable}
\def\bea{\begin{eqnarray}}
\def\eea{\end{eqnarray}}

\def\GeV{\,{\rm GeV}}

\usepackage{color}
\usepackage{ulem}

\begin{document}
\preprint{RESCEU-11/17, TU-1052, IPMU-17-0160}
\title{
Cosmological abundance of the QCD axion coupled to hidden photons
}

\author{Naoya Kitajima}
\email{kitajima.naoya@f.mbox.nagoya-u.ac.jp}
\affiliation{Asia Pacific Center for Theoretical Physics, Pohang 37673, Korea}
\affiliation{Department of Physics, Nagoya University, Nagoya 464-8602, Japan}
\author{Toyokazu Sekiguchi}
\email{sekiguti@resceu.s.u-tokyo.ac.jp}
\affiliation{Institute for Basic Science, Center for Theoretical Physics of the Universe, \\
Daejeon 34051, South Korea }
\affiliation{Research Center for the Early Universe (RESCEU), \\
Graduate School of Science, The University of Tokyo, \\
Tokyo 113-0033, Japan}
\author{Fuminobu Takahashi}
\email{fumi@tohoku.ac.jp}
\affiliation{Department of Physics, Tohoku University, Sendai, Miyagi 980-8578, Japan}
\affiliation{Kavli IPMU (WPI), UTIAS, The University of Tokyo, Kashiwa, Chiba 277-8583, Japan}

\date{\today}

\begin{abstract}
We study the cosmological evolution of the QCD axion coupled to hidden photons. For a moderately strong coupling, the motion of the axion field leads to an explosive production of hidden photons by tachyonic instability. We use lattice simulations to evaluate the cosmological abundance of the QCD axion. In doing so, we incorporate the backreaction of the produced hidden photons on the axion dynamics, which becomes significant in the non-linear regime. We find that the axion abundance is suppressed by at most ${\cal O}(10^{2})$ for the decay constant $f_a = 10^{16}$\,GeV, compared to the case without the coupling. For a sufficiently large coupling, 
the motion of the QCD axion becomes strongly damped, and as a result, the axion abundance is enhanced. 
Our results show that the cosmological upper bound on the axion decay constant can be relaxed by a few hundred for a certain
range of the coupling to hidden photons.
\end{abstract}


\maketitle

{\it Introduction} -- 
The axion is a pseudo Nambu-Goldstone boson in the Peccei-Quinn (PQ) 
mechanism\,\cite{Peccei:1977ur,Peccei:1977hh,Weinberg:1977ma,Wilczek:1977pj},
which provides a dynamical solution to the strong CP problem in the quantum chromodynamics (QCD). 
The axion is copiously  produced in the early Universe, and it is one of the plausible candidates for dark matter (DM). There are many on-going and planned 
experiments aiming to detect the axion. See Refs.~\cite{Kim:2008hd,Wantz:2009it,Ringwald:2012hr,Kawasaki:2013ae}  
for recent reviews on the QCD axion and axion-like particles.

The axion mass and coupling strengths are characterized 
by the decay constant, $f_a$. The so-called axion window between the astrophysical and cosmological 
bounds is given by
$4 \times 10^8{\rm\,GeV} \lesssim f_a \lesssim 10^{12}\,{\rm GeV}$,
where the lower bound is due to the neutrino burst duration of SN1987A~\cite{Mayle:1987as,Raffelt:1987yt,Turner:1987by}
while the upper bound is due to the cosmological abundance of the axion 
 produced by the misalignment mechanism~\cite{Preskill:1982cy,Abbott:1982af,Dine:1982ah}. 
The latter assumes the initial displacement  of order $f_a$ from the CP-conserving minimum. 
The upper bound can be relaxed if one consideres
 entropy production in the early 
Universe~\cite{Dine:1982ah,Steinhardt:1983ia,Lazarides:1990xp,Kawasaki:1995vt,Kawasaki:2004rx}, 
non-standard couplings to hidden monopoles~\cite{Kawasaki:2015lpf,Kawasaki:2017xwt,Houston:2017kwe} and a resonant mixing with 
axion-like particles~\cite{Kitajima:2014xla}.

Let us introduce the following coupling to massless hidden photons\footnote{
Such coupling of the QCD axion to hidden photons has been studied 
in Refs.~\cite{Kawasaki:2015lpf, Nomura:2015xil,Higaki:2016yqk,Ejlli:2016asd,Kaneta:2016wvf,Kawasaki:2017xwt}.
},
\bea
-\frac{g_{\phi \gamma'}}{4}\phi F_{{\rm H} \mu\nu}\tilde F_{\rm H}^{\mu\nu}
\equiv -\frac{\beta_{\rm H}}{4f_a}\phi F_{{\rm H} \mu\nu}\tilde F_{\rm H}^{\mu\nu},
\label{eq:Lint}
\eea
where $\phi$ is the axion field, $F_{{\rm H}\mu\nu}=\nabla_\mu A_{{\rm H}\nu}-\nabla_\nu A_{{\rm H}\mu}$ is the field strength of
the hidden U(1)$_{\rm H}$ gauge field $A_{{\rm H} \mu}$ and $\tilde F_{\rm H}^{\mu\nu}=\epsilon^{\mu\nu\rho\sigma}F_{{\rm H} \rho\sigma}/{2\sqrt{-g}}$
is its dual.
 Here and in what follows, we assume that there are no light hidden matter
 fields charged under U(1)$_{\rm H}$.
The coupling constant $\beta_{\rm H} \equiv g_{\phi \gamma'} f_a$ is 
model-dependent, and as pointed out in Ref.~\cite{Higaki:2016yqk} it can be much larger than unity 
if one considers the clockwork/aligned QCD axion model \cite{Higaki:2014qua,Higaki:2015jag}  (See also \cite{Kim:2004rp,Choi:2014rja,Choi:2015fiu,
Kaplan:2015fuy,Giudice:2016yja,Farina:2016tgd,Agrawal:2017cmd}).

Hidden photons are known to be explosively produced by tachyonic instability if the 
axion field evolves in time with  a moderately large coupling $\beta_H$~\cite{Garretson:1992vt}.
As a result,  the kinetic energy of the
axion efficiently turns into hidden photons. In a context of axion inflation, this explosive production of the gauge fields affects
the inflaton dynamics as well as the reheating processes~\cite{Notari:2016npn}. In particular, 
the backreaction on the axion field is significant for a large coupling, and so, it is necessary to use lattice simulation for correctly describing 
 the evolution of  the inflaton and the reheating processes~\cite{Adshead:2016iae}. 
 
The coupling of the QCD axion to hidden photons (\ref{eq:Lint}) should also modify the evolution  
of the QCD axion, if the coupling $\beta_{\rm H}$ is large enough. Recently, it was pointed out in Ref.\,\cite{Agrawal:2017eqm} that such coupling can significantly reduce the energy density of coherent oscillations of the axion. (See a note added at the end of this Letter.)  The main purpose of this Letter is to 
investigate how the coupling to hidden photons affects the cosmological abundance of the QCD axion including nonzero modes.
 Due to the explosive nature of the particle production, the system soon enters 
highly non-linear regime, which makes it difficult to describe the entire evolution analytically. We, therefore, use
lattice simulations to make a quantitative estimate of the axion abundance by incorporating the backreaction effects. 
To the best of our knowledge, this is the first lattice simulation for the QCD axion coupled
to hidden photons. 

We find that the QCD axion abundance is suppressed by at most ${\cal O}(10^{2})$
for the decay constant $f_a = 10^{16}$\,GeV for a moderately large coupling $\beta_{\rm H} \simeq 200$. 
Then, the cosmological upper bound on $f_a$ can be relaxed by at most a factor of a few hundred.
As we shall see, the reduction of the axion energy stops in a few Hubble times after the system enters non-linear regime,
and both axion and hidden photons gradually asymptote to an equilibrium state. For a sufficiently large coupling, 
on the other hand, the motion of the QCD axion becomes strongly damped, which enhances the final axion abundance compared
to the case without coupling to hidden photons. Therefore, the QCD axion abundance is not a monotonic function of
the coupling to hidden photons: 
it is suppressed for $\beta_H \lesssim 300 - 400$, while it is enhanced for $\beta_H \gtrsim 300 - 400$.

{\it Axion electrodynamics} -- 
We start from the following Lagrangian of a system with the axion $\phi$ and a hidden photon $A_{{\rm H}\mu}$:
\bea
\mathcal L&=&\frac12\nabla_\mu\phi \nabla^\mu\phi -\chi_{\rm QCD} (T)\left[1- \cos\left(\frac{\phi}{f_a}\right)\right] \notag \\
&&\quad-\frac14F_{{\rm H}\mu\nu}F_{\rm H}^{\mu\nu}-\frac{\beta_{\rm H}}{4f_a}\phi F_{{\rm H}\mu\nu}\tilde F_{\rm H}^{\mu\nu}.
\label{eq:L}
\eea
The magnitude of the axion potential is given by the QCD 
topological susceptibility $\chi_{\rm QCD}$, which is dependent on the temperature $T$.
We adopt the result from the recent lattice QCD calculation in Ref.\,\cite{Borsanyi:2016ksw} 
and approximate $\chi_{\rm QCD}(T)$ as 
\begin{equation}
\chi_{\rm QCD}(T)=\frac{\chi_0}{1+(T/T_c)^b}, 
\end{equation}
with $\chi_0=(7.6\times 10^{-2}\GeV)^4$, $T_c=0.16\GeV$ and $b=8.2$.
In the rest of this Letter, we normalize the scale factor $a$ to unity when $T=T_c$.

In the temporal gauge, $A_{{\rm H}\mu}=(0,\bm{A}_{\rm H})$, the Lagrangian \eqref{eq:L} leads 
to the following equations of motion:
\begin{align}
&\ddot\phi+2\mathcal H\dot\phi-\nabla^2 \phi+a^2 \frac{\partial V}{\partial \phi} 
=-\frac{\beta_{\rm H}}{f_a a^2} \dot{\bm{A}}_{\rm H}\cdot(\nabla\times \bm{A}_{\rm H}),
\label{eq:phi} \\
&\ddot{\bm{A}}_{\rm H}
+\nabla\times(\nabla\times\bm{A}_{\rm H}) %-\nabla^2\bm{A}_{\rm H}+\nabla(\nabla\cdot\bm{A}_{\rm H})
=\frac{\beta_{\rm H}}{f_a}\left[\dot{\phi}(\nabla\times\bm{A}_{\rm H})-(\nabla\phi)\times\dot{\bm{A}}_{\rm H}\right], 
\label{eq:A}\\
&\nabla\cdot\dot{\bm{A}}_{\rm H} =\frac{\beta_{\rm H}}{f_a}(\nabla\phi)\cdot(\nabla\times\bm{A}_{\rm H}), 
\label{eq:const} 
\end{align}
where the overdots represent derivatives with respect to the conformal time $\tau$ and
$\mathcal H=\dot a/a$ is the conformal Hubble parameter. 
Note that the last equation is the constraint equation since the longitudinal 
component of the gauge field is not dynamical as in the usual electrodynamics.

When the spatially homogeneous $\phi$ evolves in time, 
one can recast Eq.\,\eqref{eq:A} into
\bea
\ddot A_{{\rm H} \bm k,\pm}+k(k\mp\frac{\beta_{\rm H}\dot{\phi}}{f_a})A_{{\rm H} \bm k,\pm}=0
\label{eq:tachyonic}
\eea
in the Fourier space~\cite{Garretson:1992vt}, 
where $k \equiv |{\bm k}|$ is the wave number, and the subscript $\pm$ indicates the helicity.
One can see that one of the helicity components of the hidden photon
${A}_{{\rm H} \bm k,+}$ or ${A}_{{\rm H} \bm k,-}$  satisfying $k<\beta_{\rm H}|\dot{\phi}|/f_a$
becomes tachyonic.  
Thus the kinetic energy of the axion zero mode is efficiently transferred to hidden photons by the tachyonic instability.

Once a large number of hidden photons are produced by tachyonic instabilities, 
they source the axion fluctuations through Eq.~\eqref{eq:phi}. Due to their large occupation number, such reflux can 
be so efficient that the axion zero mode is soon swamped by the nonzero modes. We have solved the equations of 
motion for the axion and the gauge field and confirmed that this is indeed the case. 
To make a quantitative estimate of the axion relic density, therefore, one must resort to lattice calculations.

{\it Simulation} -- 
We have performed a simulation of the axion electrodynamics with a finite-difference method.
Specifically, we have extended the Yee's algorithm\,\cite{Yee} to accommodate the axion in the staggered grids,
and adopted a periodic comoving box with a grid number $N_{\rm grid}=256^3$. 
The time integration is implemented by the {\it leap-frog} %symplectic 
method, so that our algorithm is accurate up to the second order in both the time and spatial domains. 
The staggered grids are advantageous because the discretized constraint 
Eq.\,\eqref{eq:const} can be satisfied automatically apart from round-off errors, 
which we have explicitly verified by monitoring the gradient of $\dot{\bm{A}}_{\rm H}-\frac{\beta_{\rm H}}{f_a}\phi\nabla\times \bm{A}_{\rm H}$.
Details of our simulation will be presented in a separate paper \cite{Kitajima}.

\begin{figure*}[t]
\centering
\begin{tabular}{cc}
\includegraphics[scale=0.7]{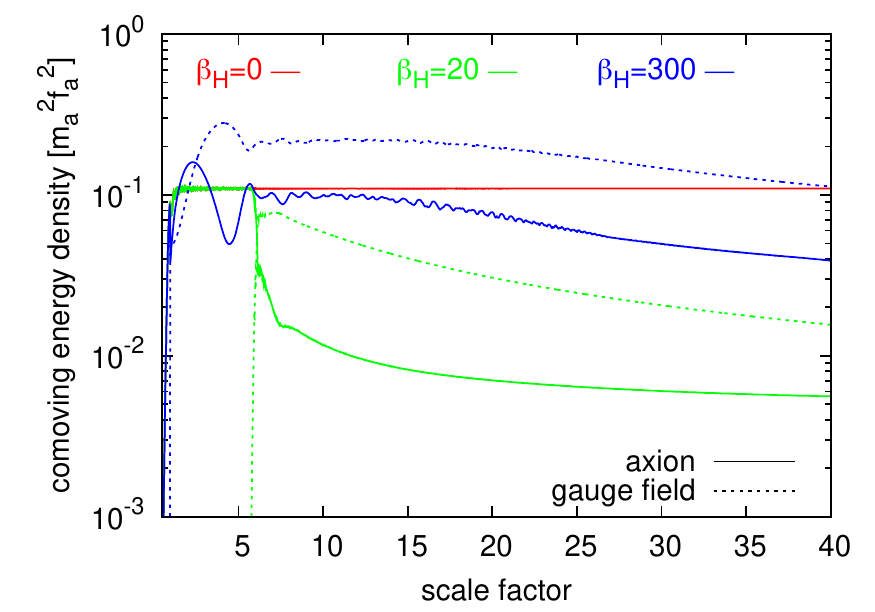} &
\includegraphics[scale=0.7]{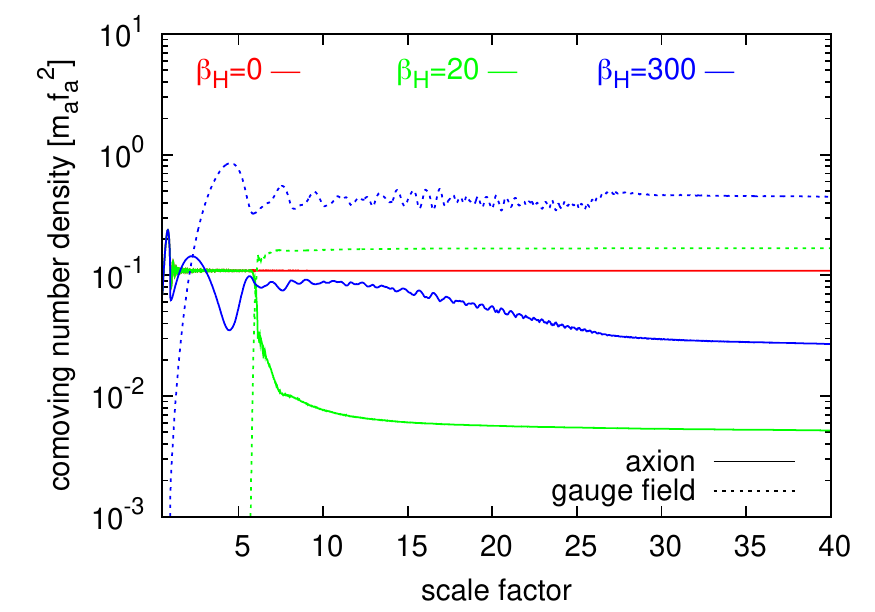}
\end{tabular}
\caption{\label{fig:evol} The evolution of the comoving energy densities (left) and the comoving number densities (right) of the axion (solid lines) and the gauge field (dotted lines).
We here adopted $\beta_{\rm H} = 0$ (red), 20 (green) and 300 (blue).
}
\end{figure*}

\begin{figure}[t]
\centering
\includegraphics[scale=0.7]{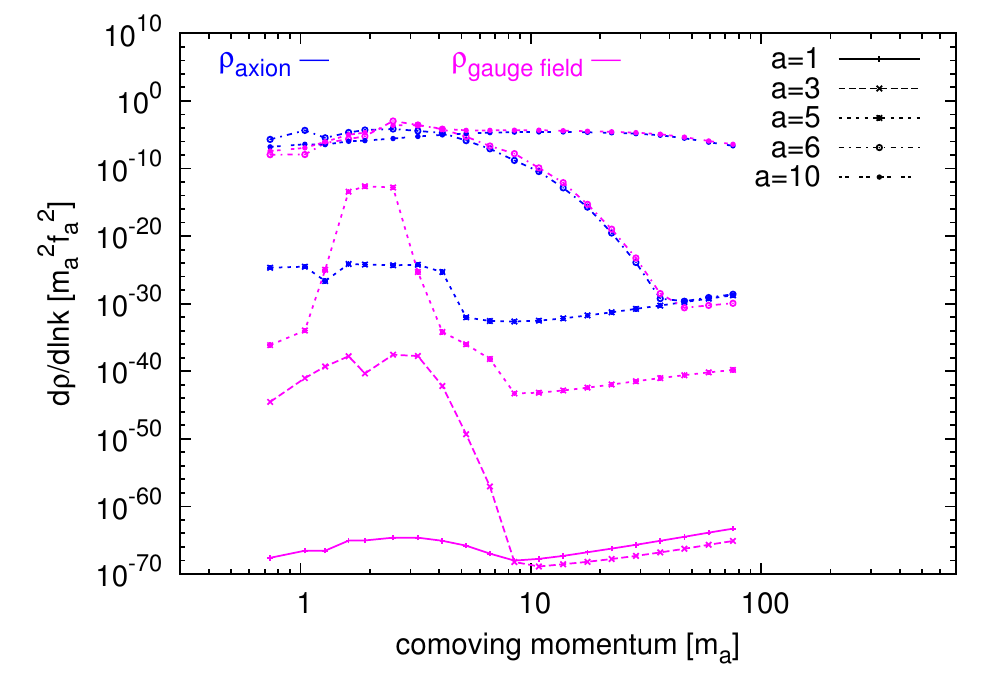}
\caption{\label{fig:spectrum} The energy spectra of the axion (blue) and the gauge field (magenta)
with $\beta_{\rm H} = 20$.
We plotted the energy spectra
at the scale factors %of 
ranging from $a=1$ to $a=10$.
}
\end{figure}

{\it Axion abundance} -- 
We show in Fig.\,\ref{fig:evol} the time-evolution of the energy densities (left) and number densities (right) of the axion and gauge field obtained in the lattice simulation
with $f_a=10^{16}\GeV$, $\beta_{\rm H}=0$ (red), 20 (green), 300 (blue) and the initial amplitude of the axion field $\phi_{\rm ini}=(\pi/3)f_a$.
One can immediately see that for $\beta_{\rm H}=20$, the axion number density gets suppressed as soon as the energy density of the gauge field becomes comparable to that of the axion. In this phase, the axion coherent oscillation is rapidly dissipated into the gauge field.
At almost the same time, the axion nonzero modes are  produced by the backreaction from the gauge field (see Fig.\,\ref{fig:spectrum}).
After this event,  the initial oscillation energy of axion is exhausted dominantly by the gauge field, 
and also fractionally by the axion nonzero modes.

For $\beta_{\rm H} = 300$ on the other hand, the gauge field increases more rapidly soon after the onset of the axion oscillation. In that case, 
the backreaction becomes significant before the 
axion completes a single oscillation and acts as a friction on the axion dynamics.
As a result, the axion rolls down the potential more slowly, which
effectively delays the onset of the axion oscillation. 
This implies that when $\beta_{\rm H}$ is further increased,
the final abundance of the axion can be enhanced rather than suppressed, 
which will be shown explicitly later.
%This potentially enhances the final abundance of the axion for larger $\beta_{\rm H}$. 
The friction dominated regime is often applied to the axion warm inflation~\cite{Notari:2016npn,Adshead:2016iae}
(See also Refs.\,\cite{Hook:2016mqo,Choi:2016kke} for a similar effect in the relaxion model).

Fig.\,\ref{fig:spectrum} shows the energy spectra of
the axion and 
gauge field for $\beta_{\rm H}=20$ plotted at several different scale factors. We can see that, at $a \lesssim 5$, 
the gauge field at low wave number is enhanced via the tachyonic instability. Subsequently, 
the axion nonzero modes are produced from the mode-mode coupling of the gauge field.
Notably, the axion energy spectrum is flat in contrast to that of the gauge field
(See the energy spectra at $a=5$). This means that even the axion nonzero modes with 
momenta higher than the maximum tachyonic wave number, i.e. $k\gg \beta_{\rm H} \dot\phi/f_a$, 
are produced. This is possible because these particles are produced from the 
mode-mode coupling of the gauge field between enhanced tachyonic mode (low momenta)
and the vacuum fluctuations (high momenta). The mode-mode coupling of the axion nonzero modes
and the gauge field sources the gauge field with different momentum as seen in the RHS of Eq.\,\eqref{eq:A}.
When $a \gtrsim 6$,  both the gauge field and the axion are populated over
a wide range of momenta with a similar spectrum. This suggests that the axion 
nonzero modes and the gauge field are coupled with each other effectively in the nonlinear regime. 

\begin{figure}[!t]
\centering
\includegraphics[scale=0.75]{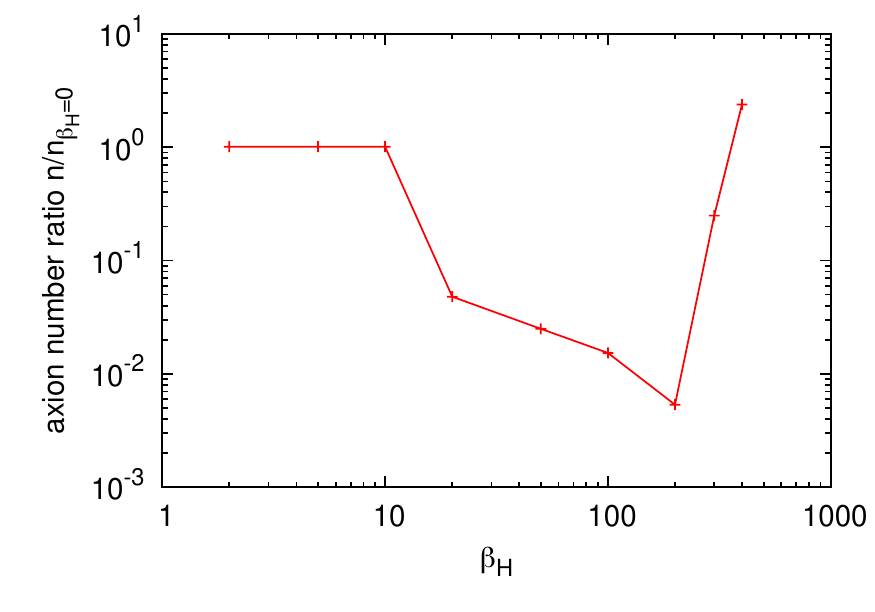}
\caption{\label{fig:ratio}
The asymptotic abundance of the axion 
normalized with the abundance of the axion in the absence of the coupling (i.e. $\beta_{\rm H}=0$).
Here we adopted $f_a=10^{16}\GeV$.
}
\end{figure}

We finally examine how the suppression of the axion abundance depends on the coupling strength $\beta_{\rm H}$.
Fig.\,\ref{fig:ratio} shows the ratio of the asymptotic abundance of the axion 
to that of the axion zero mode in the absence of the coupling. We can see from the figure that for small $\beta_{\rm H} \lesssim 10$, 
the suppression is virtually absent. This is because the coupling is not strong enough for 
 the tachyonic growth of the gauge field to affect the axion dynamics.
With moderately large $\beta_{\rm H} \gtrsim 10$, 
a large amount of the gauge field is produced, and  the axion abundance starts to be suppressed. 
As $\beta_{\rm H}$ increases,  the extent of the 
suppression of the axion abundance becomes more remarkable. 
For   $\beta_{\rm H} \sim 200$, however,  one can see that the axion abundance turns upward and
continues to increase for a larger coupling. This regime corresponds to the overdamping of the axion coherent oscillation.
We have found that it takes a longer time for the system to settle down in equilibrium for larger $\beta_{\rm H}$, and
that the final energy spectra become relatively hard. We have confirmed that our results are robust against changing
lattice resolution, simulated time, and the initial condition. Therefore, we believe that our finding that the axion abundance
is enhanced in the overdamping regime is valid, even though the enhancement factor could be slightly modified by more detailed 
lattice calculations.

The critical value of $\beta_{\rm H}$ corresponding to the turnaround between the suppression and the overdamping 
regime can be analytically estimated as follows. 
Let us focus on the brief period after the axion starts rolling but before it reaches the potential minimum.
The back reaction on the axion dynamics becomes important when
the energy density of the hidden gauge field becomes comparable to the axion energy density.
If this happens before the axion reaches the potential minimum, the system is considered to enter 
the overdamping regime.
From Eq.\,(\ref{eq:tachyonic}), one obtains the peak wavenumber of the produced hidden photons by tachyonic instability, $k_{\rm peak} \sim \beta_{\rm H} \dot\phi/2 f_a$. Near the peak wavenumber, one of the circular polarization modes of the hidden gauge field is exponentially amplified as 
\begin{equation}
A_{\rm H} \propto \exp \left( \int d\tau k_{\rm peak} \right) \sim \exp\left( \frac{\beta_{\rm H} \Delta \phi}{2f_a} \right),
\label{eq:amp}
\end{equation}
where $\Delta \phi$ is the axion field excursion during the amplification. 
Thus, the initial energy density of the hidden gauge field at the onset of the axion oscillation, 
$\rho_{\rm H,ini} \sim (k_{\rm peak}/a)^4$
% \sim (\beta_{\rm H} m_a \phi_{\rm ini}/2f_a)^4$
 is amplified by a factor of $ \exp\left(\beta_{\rm H} \Delta \phi/f_a \right)$.
Equating the energy density of the axion and hidden gauge field when the axion reaches the potential minimum,
i.e., $\Delta \phi = \phi_{\rm ini}$,
we obtain
%The back reaction on the axion dynamics becomes important when
%the energy density of the hidden gauge field becomes comparable to the axion energy density, 
%roughly given by $ \rho_{\rm H} \sim \frac{1}{2} m_a^2 \phi_{\rm ini}^2$.
%This happens when
\begin{equation}
\label{eq:Deltaphi}
\frac{1}{2} m_a^2 \phi_{\rm ini}^2 \sim \left(\frac{\beta_{\rm H} m_a \phi_{\rm ini}}{2f_a} \right)^4 \exp \left( \frac{\beta_{\rm H} \Delta\phi}{f_a} \right).
\end{equation}
%If the above relation is satisfied before the axion reaches the potential minimum, the system is considered to enter the overdamping regime.
%%the timescale of the tachyonic growth is sufficiently shorter than the oscillation timescale of the axion, 
%%the axion follows the overdamping regime. 
%It corresponds to $\Delta\phi \lesssim \phi_{\rm ini}$ and one obtains from Eq.\,(\ref{eq:Deltaphi}), 
Thus we arrive at
\begin{equation}
%\beta_H \gtrsim 4 \theta_{\rm ini}^{-1} \log \left(\frac{2 f_a}{\theta_{\rm ini} \Lambda_{\rm QCD}} \right),
\beta_{\rm H}  \sim 2 \theta_{\rm ini}^{-1} \log\left(\frac{\phi_{\rm ini}}{m_a}\right),
\end{equation}
up to a logarithmic correction, where $\theta_{\rm ini} = \phi_{\rm ini}/f_a$ is the initial misalignment angle.
For our reference values, $\theta_{\rm ini} = \pi/3$ and $f_a = 10^{16}$ GeV,
we obtain $\beta_{\rm H} \sim 200$.
%It reads $\beta_{\rm H} \gtrsim 200$ for $\phi_{\rm ini} = f_a = 10^{16}$ GeV and $m_a = \Lambda_{\rm QCD}^2/f_a$ with $\Lambda_{\rm QCD} = 0.1$ GeV.
This agrees well with our numerical result, as one can see the turnaround around $\beta_H = 200$ in Fig.~\ref{fig:ratio}.

{\it Discussion and conclusions} -- 
We have studied the evolution of the QCD axion coupled to hidden photons using lattice
calculations, and found that the backreaction of the produced hidden photons on
nonzero modes of the axion becomes significant in the non-linear regime. In particular, while
the axion density is suppressed for a moderately large coupling, it is enhanced for
a sufficiently large coupling. 
The suppression factor of the axion density can be as large as ${\cal O}(10^2)$
for $f_a = 10^{16}{\rm\,GeV}$ and $\beta_{\rm H} = 200$. 
Although the precise value of the maximum suppression factor may be altered depending on $f_a$ because it changes the relative
time scale of axion oscillation compared to the Hubble time and hence the spectral peak of the axion nonzero modes,
we believe that our main result should hold qualitatively for smaller values of $f_a$ as well as for 
axion-like particles coupled to hidden photons. In particular, one may be able to
enhance the QCD axion abundance to explain the DM abundance even for $f_a$ near the smaller 
end of the axion window for a sufficiently large $\beta_{\rm H}$.
More detailed analysis for varying $f_a$ and initial misalignment angle is left for the future work.

If $U(1)_{\rm PQ}$ symmetry gets
broken after inflation, cosmic strings and domain walls are produced.
They may experience an additional frictional force in the presence of a strong
coupling to hidden photons, which modifies the resultant axion abundance produced 
from these defects (see e.g. Ref. \cite{Kawasaki:2014sqa}).
Also, the anharmonic effect as well as axionic isocurvature perturbation 
may also be affected. We leave these issues for the future work.

{\it Note added:}
While preparing this Letter, we found Ref.\,\cite{Agrawal:2017eqm} in which its authors studied  the same set up as in Eq.\,\eqref{eq:L}. They presumed that
the backreaction effect was small for their choice of the parameters, and found that
the axion abundance can be significantly suppressed by a factor between $10^{4}$ and $10^{13}$ for $\beta_{\rm H} = 20$\,--\,30. 
Although our lattice simulations did not confirm such a large suppression factor, the effect on the axion density becomes important for a similar value of $\beta_{\rm H} \gtrsim 10$.
Also, for the limited 
suppression factor found in our calculations, the initial energy density of the QCD axion must be
 much smaller than the total energy density at the onset of oscillations. Thus, the produced hidden photons 
cannot give an appreciable contribution to the effective neutrino species.

{\it Acknowledgements} -- 
This work is supported by Grant-in-Aid for JSPS Fellows (N.K.), JSPS KAKENHI Grant Numbers 
JP15H02082 (T.S.), JP15H05889 (F.T.), JP15K21733 (F.T.), 
JP26247042 (F.T),  JP17H02875 (F.T.), JP17H02878(F.T.), and
by World Premier International Research Center Initiative (WPI Initiative), MEXT, Japan (F.T.).

%\bibliography{axion}% Produces the bibliography via BibTeX.

\onecolumngrid
\appendix
\newpage

\end{document}